# Gender-Dependent Emotion Recognition Based on HMMs and SPHMMs

Ismail Shahin

Electrical and Computer Engineering Department

University of Sharjah

P. O. Box 27272

Sharjah, United Arab Emirates

E-mail: ismail@sharjah.ac.ae

**Abstract**

It is well known that emotion recognition performance is not ideal. The work of this research is devoted to improving emotion recognition performance by employing a two-stage recognizer that combines and integrates gender recognizer and emotion recognizer into one system. Hidden Markov Models (HMMs) and Suprasegmental Hidden Markov Models (SPHMMs) have been used as classifiers in the two-stage recognizer. This recognizer has been tested on two distinct and separate emotional speech databases. The first database is our collected database and the second one is the Emotional Prosody Speech and Transcripts database. Six basic emotions including the neutral state have been used in each database. Our results show that emotion recognition performance based on the two-stage approach (gender-dependent emotion recognizer) has been significantly improved compared to that based on emotion recognizer without gender information and emotion recognizer with correct gender information by an average of 11% and 5%, respectively. This work shows that the highest emotion identification performance takes place when the classifiers are completely biased towards suprasegmental models and no impact of acoustic models. The results achieved based on the two-stage framework fall within 2.28% of those obtained in subjective assessment by human judges.

## 1. Motivation and Literature Review

Emotion recognition field is gaining an increasing momentum and attention these days because of the importance and the practicality of this field. Emotion recognition has many applications that appear in telecommunications, human robotic interfaces, smart call centers, and intelligent spoken tutoring systems [1], [2], [3]. In telecommunications, emotion recognition can be used to evaluate a caller's emotional state for telephone response services. Emotion recognition can also be used in human robotic interfaces where robots can be taught to interact with humans and recognize human emotions. More applications of emotion recognition from speech emerge in smart call centers where emotion recognition can help to spot possible problems emerging from an unsatisfactory interaction. In intelligent spoken tutoring systems, emotion recognition can be utilized to detect and adapt to students' emotions when students reach a boring state during the tutoring session.

Emotion recognition field has been studied in many occasions. The authors of [4] shed the light on recognizing emotions from spoken language. They used a combination of three sources of information for emotion recognition. The three sources are acoustic, lexical, and discourse [4]. The authors of [5] targeted in one of their works to improve the automatic emotional speech classification methods using ensemble or multi-classifier system (MCS) approaches. They also targeted to examine the differences in perceiving emotion in human speech that is obtained from different methods of acquisition [5]. In one of their studies, the authors of [6] proposed a text-independent method of emotion classification of speech based on hidden Markov models (HMMs). The authors of [7] proposed a new feature vector that assists in enhancing the classification performance of emotional states of humans. The elements of such a feature vector are attained from a feature subset selection method based on genetic algorithm [7]. In one of their works, the authors of [8] preceded their emotion recognizer by a gender recognizer to enhance emotion recognition performance.

In literature, different techniques, algorithms and models have been employed to classify emotions through speech. HMMs have been used by the authors of [6], [8], [9]. Neural Networks (NNs) have been applied by the authors of [10], [11]. Genetic Algorithms (GAs) have been exploited by the authors of [7]. Support Vector Machines (SVMs) have been implemented by the authors of [12], [13]. Suprasegmental Hidden Markov Models (SPHMMs) have been developed, used and evaluated by the author of [14] to recognize the emotional state of a speaker [15], [16].

The contribution of this work is focused on enhancing emotion identification performance based on a two-stage recognizer that is composed of gender recognizer followed by an emotion recognizer. Our system in this work is a text-independent and speaker-independent emotion recognizer. Both HMMs and SPHMMs have been used as classifiers in the two-stage architecture. This architecture has been evaluated on two different and separate emotional speech databases. The two databases are our collected database and Emotional Prosody Speech and Transcripts database. Six basic emotions including the neutral state have been used in each database. Our work in this research is different from the work in [8] in many aspects. First, our work uses both HMMs and SPHMMs as classifiers, whereas the work in [8] uses Naïve Bayes classifier. Second, we test the two-stage approach using two different English speech databases, whereas the work in [8] tests the same approach using two distinct German speech databases. Third, three extensive experiments have been carried out in our current work. Fourth, some statistical significance tests have been conducted in the present work.

The current research is different from our three previous research [15], [16], [17]. In [15], we enhanced speaker identification performance in an emotional talking environment based on using emotion cues by proposing a two-stage recognizer that combines and integrates both emotion recognizer and speaker recognizer into one recognizer. The proposed approach employs both HMMs and SPHMMs as classifiers. In [16], we studied, investigated, and analyzed emotion identification in completely two distinct and separate emotional speech databases based on SPHMMs. One database is unbiased towards any emotion, while the second one is biased towards different emotional states. In [17], we focused on improving speaker identification performance in a shouted talking environment based on a gender-dependent approach using SPHMMs.

The importance of emotion recognition clearly appears in speaker recognition. Emotion recognition has been used in enhancing speaker recognition performance in emotional talking environments based on a two-stage speaker recognition approach [15]. The author of [15] proposed a new two-stage speaker recognizer that combines and integrates both emotion recognizer and speaker recognizer into one recognizer. His results show that average speaker identification performance based on the two-stage recognizer is 79.92% with a significant enhancement over a one-stage recognizer with an identification performance of 71.58% [15].

The structure of this paper is as follows. Next section outlines the essentials of SPHMMs. Section 3 describes the two speech databases used in this work and the extraction of features. Emotion identification approach and the experiments are explained in Section 4. Discussion of the results obtained in the current work is given in Section 5. Section 6 concludes the work of this research with some remarks.

**2. Essentials of Suprasegmental Hidden Markov Models**

SPHMMs were proposed, implemented, and tested in many studies [14-18]. These models proved to be superior to HMMs for each of emotion recognition and speaker recognition in each of stressful and emotional talking environments [14-18]. The reason of this superiority is that suprasegmental state possesses the ability to look at the observation sequence through a larger window by allowing observations at rates suitable for the situation of modeling. Suprasegmental state is a state that summarizes several conventional hidden Markov states into a new state called suprasegmental state. The prosodic features of a unit of speech are characterized as suprasegmental features because they have influence on all segments of the unit of speech. Therefore, prosodic events at the levels of phone, syllable, word, and utterance are represented using suprasegmental states; on the other hand, acoustic events are represented using conventional hidden Markov states.

Prosodic and acoustic data can be combined as given in the following equation [19],

$$log\ P\left(\lambda^v, \Psi^v | O\right) = (1-\alpha).\ log\ P\left(\lambda^v | O\right) + \alpha.\ log\ P\left(\Psi^v | O\right) \tag{1}$$

where $\alpha$ is a weighting factor. Where,

$$\begin{cases} 0.5 > \alpha > 0 & \text{biased towards acoustic model} \\ 1 > \alpha > 0.5 & \text{biased towards prosodic model} \\ \alpha = 0 & \text{biased completely towards acoustic model and no effect of prosodic model} \\ \alpha = 0.5 & \text{not biased towards any model} \\ \alpha = 1 & \text{biased completely towards prosodic model and no impact of acoustic model} \end{cases} \tag{2}$$

$\lambda^v$ is the acoustic model of the $v^{th}$ emotion, $\Psi^v$ is the suprasegmental model of the $v^{th}$ emotion, $O$ is the observation vector or sequence of an utterance, $P\left(\lambda^v | O\right)$ is the probability of the $v^{th}$ HMM emotion model given the observation vector $O$, and $P\left(\Psi^v | O\right)$ is the probability of the $v^{th}$ SPHMM emotion model given the observation vector $O$. More information about Suprasegmental Hidden Markov Models can be found in [14-18].

## 3. Speech Databases and the Extraction of Features

Two distinct emotional speech databases of English language have been used in this work to test the two-stage emotion recognizer. The two databases are: our collected database and Emotional Prosody Speech and Transcripts database.

**3.1 The Collected Database**

The collected speech database was captured from thirty (fifteen male and fifteen female) untrained healthy adult native speakers of American English. Untrained speakers (uttering sentences naturally) were chosen to avoid overstressed expressions. Each speaker uttered eight sentences where each sentence was portrayed five times in one session (training session) and four times in another separate session (test session) under each of the neutral, angry, sad, happy, disgust, and fear emotions. The eight sentences were unbiased towards any emotion when uttered under the neutral state. These sentences are:

1) *He works five days a week.*
2) *The sun is shining.*
3) *The weather is fair.*
4) *The students study hard.*
5) *Assistant professors are looking for promotion.*
6) *University of Sharjah.*
7) *Electrical and Computer Engineering Department.*
8) *He has two sons and two daughters.*

This database was collected in a clean environment by a speech acquisition board using a 16-bit linear coding A/D converter and sampled at a sampling rate of 16 kHz. The database was a wideband 16-bit per sample linear data. The signal samples were segmented into frames of 16 ms each with 9 ms overlap between consecutive frames. The speech signals were applied every 5 ms to a 30 ms Hamming window.

**3.2 Emotional Prosody Speech and Transcripts Database**

Emotional Prosody database was produced by the Linguistic Data Consortium (LDC) [20]. This database is composed of eight professional speakers (three actors and five actresses) producing a series of semantically neutral utterances consist of dates and numbers spoken in fifteen different emotions. These emotions are neutral, hot anger, cold anger, panic, anxiety, despair, sadness, elation, happiness, interest, boredom, shame, pride, disgust, and contempt [20]. Only six emotions have been used in this work. The six emotions are neutral, hot anger, sadness, happiness, disgust, and panic.

**3.3 Extraction of Features**

In the current work, the features that characterize the phonetic content of speech signals in the two databases are called Mel-Frequency Cepstral Coefficients (static MFCCs) and delta Mel-Frequency Cepstral Coefficients (delta MFCCs). These coefficients have been extensively used in many studies in the fields of speech recognition [21], [22], speaker recognition [23], [24], and emotion recognition [4], [8], [13], [25]. MFCCs outperform other coefficients in the three fields and they yield a high-level approximation of human auditory perception.

MFCCs are computed with the help of a psycho acoustically motivated filter bank followed by a logarithmic compression and Discrete Cosine Transform (DCT). The following formula shows how to compute these coefficients [26],

$$C(n) = \sum_{m=1}^{M} \left\{ [log \ Y(m)] \cos\left[\frac{\pi n}{M}(m - \tfrac{1}{2})\right] \right\} \quad (3)$$

where $Y(m)$ are the outputs of an $M$-channel filter bank. In this work, a 16-dimension static and delta MFCC feature analysis (8 static and 8 delta) was used to form the observation vectors in each of HMMs and SPHMMs.

Most of the studies performed in the last few decades in each of speech recognition, speaker recognition, and emotion recognition on HMMs have been done using Left-to-Right Hidden Markov Models (LTRHMMs) because phonemes follow strictly the left-to-right sequence [6], [27], [28]. In the present work, Left-to-Right Suprasegmental Hidden Markov Models (LTRSPHMMs) have been derived from LTRHMMs. Fig. 1 illustrates an example of a fundamental structure of LTRSPHMMs that has been obtained from LTRHMMs. In this figure, $q_1$, $q_2$, ..., $q_6$ are conventional hidden Markov states. $p_1$ is a suprasegmental state that is composed of $q_1$, $q_2$ and $q_3$. $p_2$ is a suprasegmental state that is made up of $q_4$, $q_5$ and $q_6$. $p_3$ is a suprasegmental state that consists of $p_1$ and $p_2$. $a_{ij}$ is the transition probability between the $i^{th}$ conventional hidden Markov state and the $j^{th}$ conventional hidden Markov state. $b_{ij}$ is the transition probability between the $i^{th}$ suprasegmental state and the $j^{th}$ suprasegmental state.

In this work, a continuous mixture observation density is selected for the recognizer. The number of conventional states of LTRHMMs, $N$, is nine, while the number of suprasegmental states in LTRSPHMMs is three. As a result, each three conventional states of LTRHMMs are summed up into one suprasegmental state. The number of mixture components, $M$, is ten per state. The transition matrix, $A$, of such a structure can be expressed in terms of the positive coefficients $b_{ij}$ as,

$$A = \begin{bmatrix} b_{11} & b_{12} & b_{13} \\ b_{21} & b_{22} & b_{23} \\ b_{31} & b_{32} & b_{33} \end{bmatrix}$$

## 4. Emotion Identification Approach and the Experiments

This work is focused on employing a two-stage emotion recognizer that consists of two cascaded recognizers as illustrated in Fig. 2. This figure shows that the emotion recognizer is nothing but a two-stage recognizer that integrates and combines gender recognizer and emotion recognizer into one system. The two recognizers are:

**Gender Identification Recognizer**

The first stage of the two-stage architecture is to identify the gender of the speaker so as to make the output of this stage gender-dependent. Generally, automatic gender identification gives high performance with little effort because the output of this stage is the speaker either a male or female. Therefore, gender identification is a binary categorization problem which is usually simple.

Vogt and Andre preceded their emotion recognizer by a gender recognizer to enhance emotion recognition performance [8]. Lee and Narayanan demonstrated that gender-specific emotion recognition performance is higher than that when both genders are mixed [4]. Ververidis and Kotropoulos showed that the combined emotion recognition of male and female is higher than the gender-independent emotion recognition [27]. Abdulla and Kasabov improved speech recognition performance through gender separation by separating the datasets based on gender to build gender-dependent HMM for each word [29]. Their results showed significant improvement of word recognition performance based on gender-dependent method compared to the gender-independent method. Shahin focused in one of his studies on enhancing speaker identification performance in the shouted talking environment based on gender cues using SPHMMs as a classifier [17].

Two probabilities per utterance are computed in this stage based on HMMs and the maximum probability is chosen as the identified gender as given in the following formula,

$$G^* = \arg \max_{2 \geq g \geq 1} \left\{ P\left(O \mid \Gamma^g\right) \right\} \qquad (4)$$

where $G^*$ is the index of the identified gender (either $M$ or $F$), $\Gamma^g$ is the $g^{th}$ HMM gender model (one male model and one female model), and $P\left(O \mid \Gamma^g\right)$ is the probability of the observation sequence $O$ that belongs to the unknown gender given the $g^{th}$ HMM gender model.

In the training session of this stage, HMM male model has been derived using ten of the fifteen male speakers uttering all the first four sentences under all the emotions including the neutral state, whereas HMM female model has been obtained using ten of the fifteen female speakers generating all the first four sentences under all the emotions including the neutral state. Each HMM gender model has been constructed using a total of 2160 utterances (10 speakers × 4 sentences × 9 utterances/sentence × 6 emotions).

**Emotion Identification Recognizer**

Given that the gender of a speaker was identified by the previous recognizer, the aim of emotion recognizer is to identify the unknown emotional state of the speaker. This recognizer is called gender-specific emotion identifier. In this recognizer, $m$ probabilities per gender are computed based on SPHMMs. The maximum probability is selected as the identified emotion per gender as given in the following formula,

$$E^* = \arg \max_{m \geq e \geq 1} \left\{ P\left(O \mid G^*, \lambda^e, \Psi^e\right) \right\} \tag{5}$$

where $E^*$ is the index of the identified emotion, $\left(\lambda^e, \psi^e\right)$ is the $e^{\text{th}}$ SPHMM emotion model, and $P\left(O \mid G^*, \lambda^e, \Psi^e\right)$ is the probability of the observation sequence $O$ that belongs to the unknown emotion given the identified gender and the $e^{\text{th}}$ SPHMM emotion model.

In the training session, the $e^{\text{th}}$ SPHMM emotion model $(\lambda^e, \psi^e)$ per gender has been constructed for every emotion using ten (same 10 speakers in the gender identification recognizer) of the fifteen speakers uttering all the first four sentences with a repetition of nine utterances/sentence. Each SPHMM emotion model per gender has been constructed using a total of 360 utterances (10 speakers × 4 sentences × 9 utterances/sentence). The SPHMM training session is similar to the conventional HMM training session. In the SPHMM training session, suprasegmental models are trained on top of acoustic models of HMM.

In the evaluation session, each one of the remaining male speakers (unused male speakers in the training session) and the remaining female speakers (unused female speakers in the training session) used nine utterances per sentence (the last four sentences of the database that have been unused in the training session) under each emotion including the neutral state. Therefore, our system in this work is text-independent and speaker-independent. The total number of utterances used in this session is 2160 (5 speakers × 2 genders × 4 sentences × 9 utterances/sentence × 6 emotions). A block diagram of emotion identification recognizer is shown in Fig. 3.

## 5. Results and Discussion

In this work, a two-stage emotion recognizer based on both HMMs and SPHMMs has been exploited to enhance emotion identification performance. This recognizer has been evaluated using separately each of the collected and Emotional Prosody Speech and Transcripts databases when the weighting factor ($\alpha$) is equal to 0.5. This particular value of the weighting factor has been selected to avoid biasing towards any model.

Using the two-stage emotion recognizer, automatic gender identification performance based on HMMs is 95.13% and 96.77% using the collected and Emotional Prosody databases, respectively. The achieved gender identification performance in this work is higher than that reported by Vogt and Andre. They obtained gender identification performance of 90.26% and 91.85% using Berlin and SmartKom German databases, respectively [8].

Table 1 gives emotion identification performance using each of the collected and Emotional Prosody databases based on: gender-dependent emotion recognizer, emotion recognizer without gender information, and emotion recognizer with correct gender information. Using the collected database, this table yields an average emotion identification performance of 85.58%, 77.18%, and 81.99% based on gender-dependent emotion recognizer, emotion recognizer without gender information, and emotion recognizer with correct gender information, respectively. The table gives an average emotion identification performance of 86.79%, 78.06%, and 82.31% based, respectively, on gender-dependent emotion recognizer, emotion recognizer without gender information, and emotion recognizer with correct gender information using Emotional Prosody database. It is evident from this table that gender-dependent emotion recognizer outperforms the other two recognizers.

A statistical significance test has been performed to show whether emotion identification performance differences (emotion identification performance based on approach $x$ and that based on approach $y$) are real or simply due to statistical fluctuations. The statistical significance test has been carried out based on the Student's *t*-distribution test as given by the following formula,

$$t_{\text{approach x approach y}} = \frac{\overline{x}_{\text{approach x}} - \overline{x}_{\text{approach y}}}{SD_{\text{pooled}}} \qquad (6)$$

where $\bar{x}_{approach\ x}$ is the mean of the first sample (approach $x$) of size $n$, $\bar{x}_{approach\ y}$ is the mean of the second sample (approach $y$) of the same size, SD $_{pooled}$ is the pooled standard deviation of the two samples (approaches) given as,

$$SD_{pooled} = \sqrt{\frac{SD^2_{approach\ x} + SD^2_{approach\ y}}{n}} \qquad (7)$$

where SD $_{approach\ x}$ is the standard deviation of the first sample of size $n$, SD $_{approach\ y}$ is the standard deviation of the second sample of the same size.

Based on Table 1 and using the last two equations, the calculated $t$ values using each of the collected and Emotional Prosody databases between each of approach 1 and approaches 2 and 3 are given in Table 2. Each calculated $t$ value in this table is greater than the tabulated critical value at *0.05* significant level $t_{0.05}$ = 1.645. Therefore, the conclusion that can be drawn in this experiment is that emotion identification based on approach 1 outperforms that based on each of approach 2 and approach 3. Therefore, inserting gender identification recognizer into emotion identification system significantly enhances emotion identification performance compared to that without gender information and with correct gender information.

Gender-dependent emotion identification performance is better than the emotion identification performance obtained by:
  i)  Ververidis and Kotropoulos [27]. Based on their work, they achieved an average male and female emotion identification performance of 61.10% and 57.10%, respectively.
  ii) Vogt and Andre [8]. In their work, they obtained gender-dependent emotion identification performance of 86.00% and 76.70% using Berlin and SmartKom German databases, respectively. Based on their results, gender-dependent emotion identification performance has been improved compared to that of gender-independent emotion identification by 2% and 4% using Berlin and SmartKom databases, respectively.

Three extensive experiments have been conducted in this work to evaluate and verify the results achieved based on the two-stage emotion recognizer. The three experiments are:

(1) **Experiment 1**: In the emotion identification recognizer of the two-stage recognizer, the *m* probabilities were computed based on SPHMMs. To compare the impact of using acoustic features on emotion identification with that using suprasegmental features, Eq. (5) has become as,

$$E^* = \arg\max_{m \geq e \geq 1} \left\{ P\left(O \middle| G^*, \lambda^e \right) \right\} \tag{8}$$

Therefore, the *m* probabilities in this experiment are computed based on HMMs. The obtained average emotion identification performance based on the two-stage recognizer using only HMMs in both gender and emotion recognizers of the collected and Emotional Prosody databases is 80.55% and 81.17%, respectively. It is apparent from this experiment that SPHMMs are superior to HMMs for emotion identification based on the two-stage recognizer.

(2) **Experiment 2**: The two-stage emotion recognizer has been evaluated for distinct values of the weighting factor ($\alpha$). Figures 4 and 5 show average emotion identification performance based on this recognizer for different values of $\alpha$ (0.0, 0.1, …, 0.9, 1.0) using the collected and Emotional Prosody databases, respectively. The two figures show that as the value of the weighting factor increases, the average emotion identification performance (excluding for the neutral state) improves significantly. Therefore, it is evident, based on the two-stage recognizer, that SPHMMs have more impact on emotion identification performance than HMMs. The two figures also show that the highest emotion identification performance happens when $\alpha = 1$ (when the classifiers are totally biased towards suprasegmental models and no influence of acoustic models).

(3) **Experiment 3**: An informal subjective assessment of the two-stage emotion recognizer using the collected database was carried out with ten nonprofessional listeners. A total of 720 utterances (15 speakers per gender × 6 emotions × the last 4 sentences of the database) were used in this evaluation. During the evaluation, the listeners were asked to answer two questions for every test utterance. The two questions were: identify the unknown gender and identify the unknown emotion. The average gender and emotion identification performance was 96.50% and 83.67%, respectively. These averages fall within 1.44% and 2.28%, respectively, of the achieved average gender and emotion identification performance based on the two-stage recognizer using the collected database. Figure 4 demonstrates that the two-stage emotion recognizer outperforms the subjective assessment when the weighting factor is greater than 0.2.

## 6. Concluding Remarks

In this work, a two-stage emotion recognizer based on using both gender and emotion cues has been implemented and tested to improve the low emotion identification performance. The current work shows that the two-stage emotion recognizer yields higher emotion identification performance than each of emotion recognizer without gender information and emotion recognizer with correct gender information. This is because the two-stage recognizer possesses both gender and emotion cues combined and integrated into one system. It seems that the combination of both gender and emotion cues gives the superiority of the two-stage recognizer over the other two recognizers. This work also shows that SPHMMs outperform HMMs in the emotion identification stage and hence SPHMMs outperform HMMs in the overall emotion identification performance. Finally, the optimum emotion identification performance based on the two-stage approach occurs when the weighting factor is equal to one (when the classifiers are fully biased towards suprasegmental models and no influence of acoustic models).

This work carries some limitations. First, the processing computations and the time consumed in the two-stage approach are greater than those in the one-stage approach. Second, the two cascaded recognizers might perform worse than the one-stage recognizer when the emotion recognizer of the two cascaded recognizers receives false input from the gender recognizer (false identified gender) and so, errors are accumulated. Finally, the overall performance of the two-stage recognizer is limited to the resultant of two performances. The reasons of the limitations are:
   a) Gender identification recognizer does not yield ideal results. Gender identification performance is less than 100%.
   b) The unknown emotion in the emotion identification recognizer is imperfectly identified.

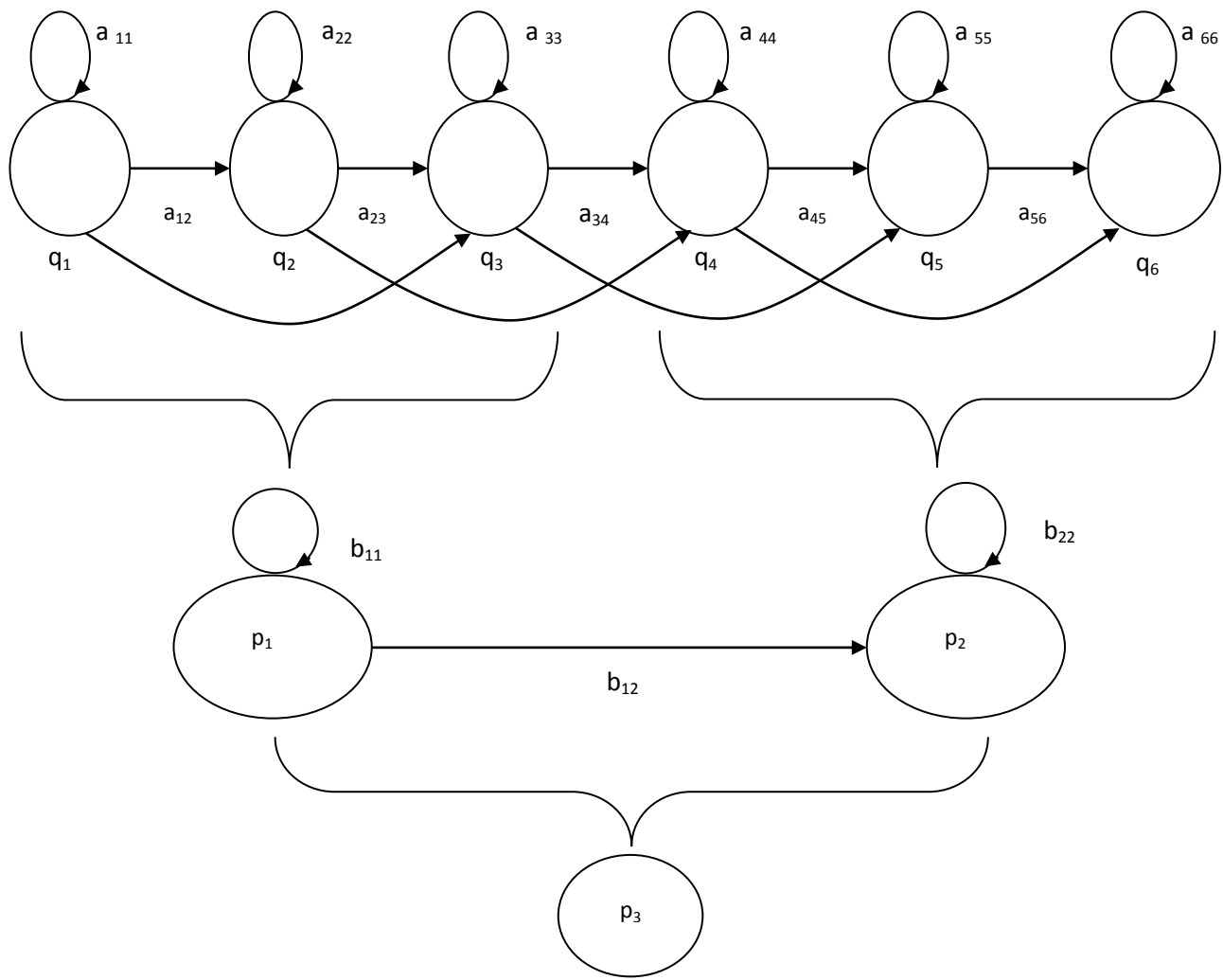

**Figure 1.** Basic structure of LTRSPHMMs

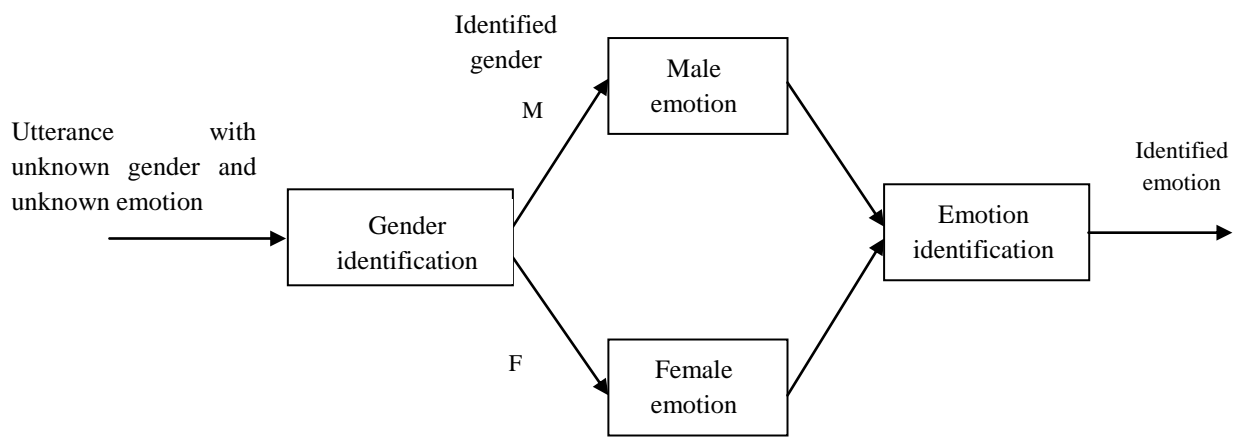

**Figure 2.** Block diagram of the overall emotion recognizer

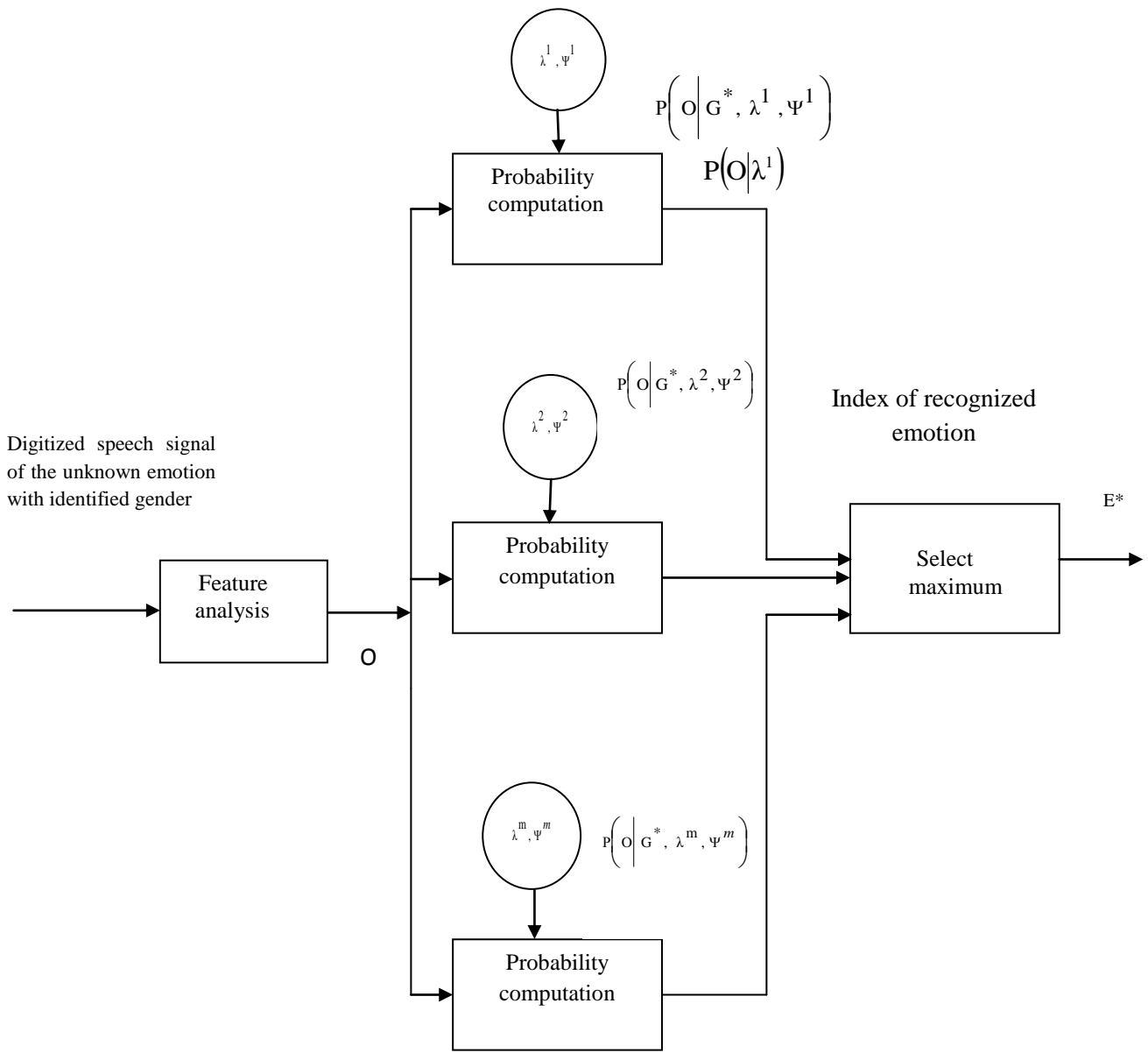

**Figure 3.** Block diagram of emotion identification recognizer

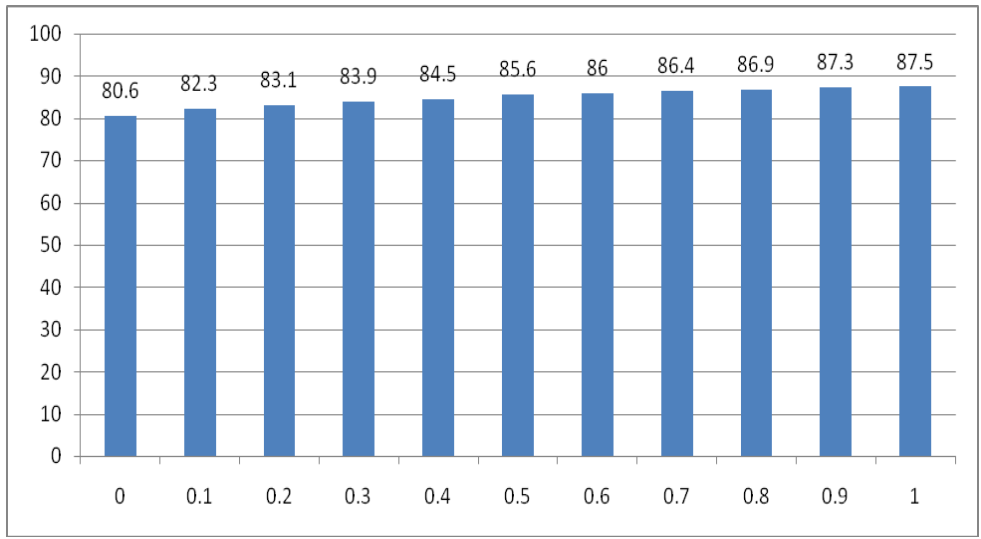

**Figure 4.** Average emotion identification performance (%) based on the two-stage emotion recognizer versus the weighting factor ($\alpha$) using the collected database

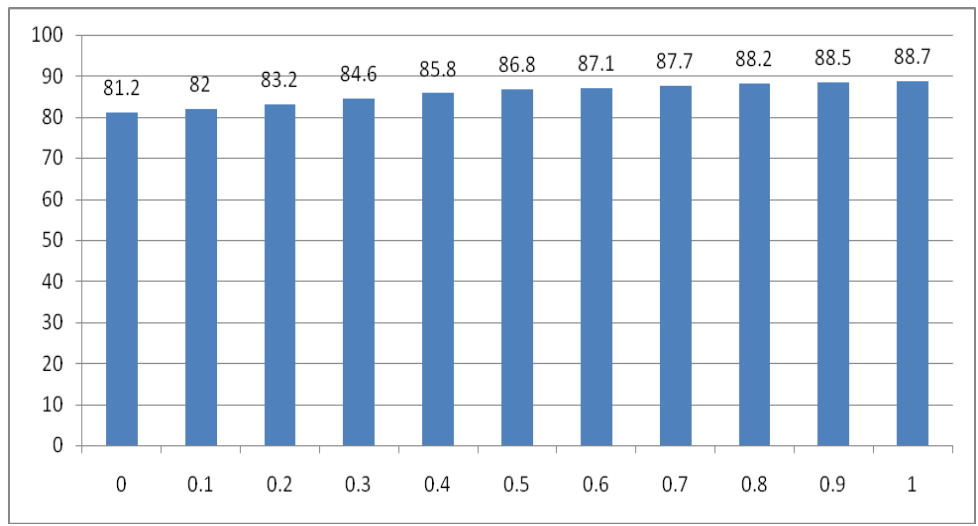

**Figure 5.** Average emotion identification performance (%) based on the two-stage emotion recognizer versus the weighting factor ($\alpha$) using the Emotional Prosody database

Table 1

Emotion identification performance using each of the collected and Emotional Prosody databases based on: gender-dependent emotion recognizer, emotion recognizer without gender information, and emotion recognizer with correct gender information

| Recognizer | Gender | Emotion identification performance (%) | |
|---|---|---|---|
| | | Collected database | Emotional Prosody database |
| Gender-dependent emotion recognizer (**approach 1**) | Male | 85.19 | 86.54 |
| | Female | 85.97 | 87.04 |
| | Average | 85.58 | 86.79 |
| Emotion recognizer without gender information (**approach 2**) | Average | 77.18 | 78.06 |
| Emotion recognizer with correct gender information (**approach 3**) | Male | 81.65 | 82.13 |
| | Female | 82.33 | 82.49 |
| | Average | 81.99 | 82.31 |

Table 2

Calculated $t$ values using each of the collected and Emotional Prosody databases between each of approach 1 and approaches 2 and 3

| Database | Calculated $t$ values | |
|---|---|---|
| Collected | $t_{\text{approach 1, approach 2}}$ | 1.985 |
| | $t_{\text{approach 1, approach 3}}$ | 1.724 |
| Emotional Prosody | $t_{\text{approach 1, approach 2}}$ | 1.886 |
| | $t_{\text{approach 1, approach 3}}$ | 1.698 |